\documentclass[prl,twocolumn,showpacs]{revtex4}
\usepackage{graphicx,bm,amssymb,amsmath,verbatim}
\usepackage{color}
\usepackage{amscd}
\usepackage{epstopdf}
\usepackage{chemarr}
\usepackage{enumerate}
\usepackage{bm}

\newcommand{\vect}[1]{\boldsymbol{#1}}

\newcommand{\dd}{\text{d}}

\def\be{\begin{equation}}
\def\fin{\end{equation}}

\def\T{{\sf T\kern-.45em T}}
\def\C{\kern.1em{\raise.47ex\hbox{$\scriptscriptstyle |$}}
       \kern-.40em{\sf C}}

\newcommand{\ds}{\displaystyle}
\newcommand{\beq}{\begin{equation}}
\newcommand{\eeq}{\end{equation}}
\newcommand{\beqa}{\begin{eqnarray}}
\newcommand{\eeqa}{\end{eqnarray}}
\newcommand{\aver}[1]{\left\langle {#1}\right\rangle}
\newcommand{\bem}{\begin{math}}
\newcommand{\eem}{\end{math}}

\allowdisplaybreaks 

\begin{document}
\title{Stress reorganisation and response in active solids}
\date{\today}

\author{Rhoda J. Hawkins$^{1,2}$\footnote{rhoda.hawkins@physics.org} and Tanniemola B. Liverpool$^2$\footnote{t.liverpool@bristol.ac.uk}}
\affiliation{$^1$Department of Physics and Astronomy, University of Sheffield, 
Sheffield S3 7RH, United Kingdom}
\affiliation{$^2$School of Mathematics, University of Bristol, Bristol BS8 1TW, United Kingdom}

\begin{abstract}
We present a microscopic model of  a disordered viscoelastic active solid, i.e. an active material whose long time behaviour is elastic as opposed to viscous. It is composed of filaments, passive cross-links and molecular motors powered by stored chemical energy, e.g. actomyosin powered by ATP. Our model allows us to study the 
collective behaviour of contractile active elements and how their interaction with each other and the passive elastic elements 
 determines the macroscopic  mechanical properties of the active material.  As a result of the (un)binding dynamics of the active elements, we find that this system provides a highly responsive material with a dynamic mechanical response strongly dependent on the {\em amount} of deformation. 
\end{abstract}

\maketitle

\paragraph{Introduction}

Active materials are condensed matter systems self-driven out of equilibrium by  
components that convert stored energy into movement. They  have generated much interest  in recent years, both as inspiration for 
a new generation of smart materials and as a framework to understand 
aspects of  cell motility~\cite{Joanny-RMP,Julicher:2007rx,Toner2005,Ramaswamy2010}.  The eukaryotic cell cytoskeleton provides a paradigmatic example of such an active material. It consists of a network of protein filaments and  associated proteins such as cross-links binding filaments together and molecular motors consuming chemical energy 
to exert forces on filaments \cite{alberts}. It shows a rich variety of behaviours including mechanical tasks involved in cell locomotion and division. 
Due to its complexity a complete physical description of the cell cytoskeleton is not currently possible. However, a fruitful direction of progress 
comes from experiments 
on simplified in-vitro systems of a small number of its components 
(namely specific filaments, cross-links and  molecular motors)~\cite{Mizuno2007,Bendix2008,Koenderink2009,Mackintosh2010,Silva2011,Kohler2011,Kohler2012}. 

The cytoskeleton is a viscoelastic material and the 
appropriateness 
 of considering it as a liquid or solid depends on timescale.
For cellular processes occurring on very long timescales, 
 large scale network remodelling due e.g. to 
 (un)binding cross-links leads to fluid-like behaviour. 
There is now a large body of theoretical work considering aspects of cytoskeletal dynamics by modelling it as an oriented active fluid, supplementing the equations of 
fluid mechanics with additional active stresses coupled to local orientational order \cite{Joanny-RMP,Liverpool2003,Kruse2004,Kruse2005,Kruse2006,Julicher:2007rx,Voituriez2005,Voituriez2006,Liverpool2006,Howard2001,Chowdhury2013,MacKintosh2008,Weber2013,FlorianThueroff2013,Simha2002,Hatwalne2004,Marenduzzo:2007,Sokolov2009,Pedley1992}. 
Although appropriate for 
many cellular processes in-vivo,
recent mechanical experiments on simplified cytoskeletal extracts are done on timescales short compared to cross-link lifetimes, when the network behaves like a disordered solid \cite{Mizuno2007,Bendix2008,Koenderink2009,Mackintosh2010,Schwarz2013,Style2013}. To understand them requires an equivalent theoretical picture of active elastic solids  \cite{MacKintosh2008,Liverpool2009,Banerjee2011,Lenz2012,Lenz2012a}. 
To address this we develop 
a microscopic model of the interactions of stress generating elements (motors) and filaments in an elastic material. 
Such models form an essential bridge between 
in-vitro and in-vivo observations, linking the macroscopic properties of the gel to the mechano-chemical properties of its components.

We first present a generic description of the linear elasticity of an active solid, highlighting the changes in mechanical properties upon switching on activity.  
We then present a microscopic stochastic model of a one dimensional disordered solid composed of both elastic and active elements, 
appropriate for describing a material on timescales for which the cross-links are fixed but the motor (un)binding dynamics are relevant. 
We investigate the collective dynamics of 
active and elastic elements, focusing on their steady state behaviour. We obtain the statistical ground state (defined as the configuration 
when no external force is applied) of the active solid and also its active elastic modulus. As expected we obtain a contractile ground state but interestingly we find that taking account of (un)binding 
 dynamics of the active elements leads to a {\em larger} contraction and a {\em smaller} elastic modulus.
This has a spectacular effect on the dynamical mechanical response: for a specific range of deformation the stress response changes sign.

\paragraph{General linear elasticity}
For small deviations, $\vect{u}$ (with $i$th component $u_i$), from a ground state, $\vect{r}_0$, the free energy, $F$, of a passive linearly elastic body is quadratic in the local strain tensor, $U_{ij}(\vect{r})=\frac{1}{2}(\partial_j u_i+\partial_i u_j)$ (where $\partial_i=\frac{\partial}{\partial x_i}$) at position $\vect{r}=(\vect{r}_0+\vect{u} (\vect{r}))$, i.e. 
\begin{math}
F=\int \dd r\, \left (\frac{1}{2}E_{ijkl}\,U_{ij}(\vect{r})U_{kl}(\vect{r}) 
\right )  ,
\end{math}
where $E_{ijkl}$ is the elastic modulus tensor.
Under an external force density, $\vect{f}(\vect{r})$, 
local force balance implies 
\begin{equation}
\partial_j \sigma_{ij}^e= \frac{1}{2}E_{ijkl}\left (\partial_l \partial_j u_k+\partial_k\partial_j u_l\right )=-f_i(\vect{r}) \; ,
\label{eq:genelforcebalance}
\end{equation}
where 
\begin{math}
\sigma_{ij}^e(\vect{r})=\frac{\delta F}{\delta U_{ij}(\vect{r})}=E_{ijkl}\,U_{kl}(\vect{r})  \; 
\end{math}
is the  local stress tensor.  
The mechanical properties of an active material, being out of equilibrium, cannot be obtained from a free energy.
Its behaviour must be constructed using dynamical arguments, respecting the conservation laws and symmetries of the system. 
For an isotropic active solid {\em in a non-equilibrium steady state}, the local force balance, eqn. (\ref{eq:genelforcebalance}) is modified by the addition of an active component, $\vect{\sigma}^a$, to the stress tensor, $\vect{\sigma}=\vect{\sigma}^e + \vect{\sigma}^a$. 
This active stress for an isotropic material, has the form
\begin{equation}
\sigma_{ij}^a=\Delta\mu\,(\zeta\,\delta_{ij}+{\cal{E}}_{ijkl}\,U_{kl}) + O (U^2),
\label{eq:genelactivestress}
\end{equation}
which are the only linear homogeneous terms allowed by symmetry. $\Delta\mu$ represents the chemical energy available in the system, derived from the chemical potential of ATP hydrolysis and $\zeta$ and ${\cal{E}}_{ijkl}$ are active parameters.
This leads to a modification of the elastic constants $\vect{E}$ to $\tilde{\vect{E}}=\vect{E}+\Delta\mu\, \vect{\cal{E}}$ and the ground state strain in the unstressed state  $\tilde{\vect{U}}_0=-\zeta\Delta\mu\,\vect{E}^{-1}\vect{I}$.
%~and $\tilde{\vect{u}}=\vect{r}-\tilde{\vect{r}}_0$.
The change in ground state strain is due to the isotropic, pressure-like term in  eqn. (\ref{eq:genelactivestress}). 
The main aim of this work is to calculate the active parameters $\zeta$ and ${\cal{E}}_{ijkl}$ from properties of the microscopic elements of the material. As a first step we consider a 1D model. 
From this model we derive the active elastic modulus, $\tilde{E}$, and ground state strain, $\tilde{U}_0$ (in 1D we use the ground state displacement $\tilde{x}_0=b_0\tilde{U_0}$).

\paragraph{Microscopic model}
\begin{figure}[hbt]
\begin{centering}
\includegraphics[width=0.48\textwidth]{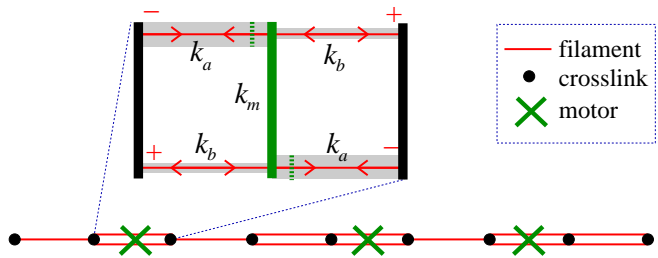}
\caption{\label{fig:1Dmuscopic}
Schematic diagram of a disordered 1D lattice consisting of single and
double bonds representing filaments (lines), cross-links (dots) and motors (crosses).
{\it {Inset:}} Cartoon of the model of an active bond discussed in section  {\em The contractile active element}. Antiparallel filaments (horizontal lines with polarity marked by $\pm$) are linearly elastic with spring constants $k_{a,b}$ and arrows show filament sections under compression/extension. The motor unit (middle vertical line) can also be elastic with $k_m$. Dashed lines indicate initial position of the motor before stepping and extension/compression.
}
\end{centering}
\end{figure}
We consider a 1D lattice of $N$ sites representing passive cross-links, connected by bonds representing filaments, as depicted in Fig.~\ref{fig:1Dmuscopic}~(bottom). 
Each site has one associated bond, labelled by $i=1\dots N$, of  
 length $b_i$ and unstretched length $b_0$ i.e. $x_i=b_i-b_0$ is the extension of bond $i$. The unstretched system length is
 $L_0 = N b_0$ and undeformed volume $V=L_0 A$ where $A$ is the cross-sectional area.
 There are $N_f$ filaments distributed on the bonds giving filament density as $\rho=N_f/N$.
For mechanical stability in 1D, all bonds must be present to have a percolated system.
Respecting this but allowing some disorder, we allow bonds to be either single, with fraction $\phi_1$, or double, with fraction $\phi_2=1-\phi_1=\rho-1$. 
Assuming each filament is placed randomly the fraction of bonds with multiplicity $r$ is given by $\phi_r=\frac{(\rho N)!}{r!(\rho N-r)!}\left(\frac{1}{N}\right)^r\left(\frac{N-1}{N}\right)^{\rho N-r}$. For our assumption of only single and double bonds to be valid with a $20\%$ tolerance, the range of acceptible number densities is $1.6<\rho<2.2$~\footnote{We should note that in the model described below, in the case of a large number of active bonds, there is the unphysical possibility of having more
active bonds than there are double bonds. In the case of all double bonds, the distribution tail is
correct, however for $\rho<2$ the distribution tail will not be correct.
}.

The filament polarity $p$ is defined as the fraction $p_R$ of bonds
pointing right minus the fraction pointing left; $p=p_R-(1-p_R)$.
The number density of bound motors 
is given by $m=(\text{number of motors})/N$. Only antiparallel double bonds with a motor bound are active~\cite{Liverpool2009,Lenz2012}. The
fraction $a$ of active bonds is therefore given by $a=2\,p_R(1-p_R)m\,\phi_2=\frac{1}{2}(1-p^2)m(\rho-1)$.
The architecture of the lattice sites and bonds is taken as constant, $\rho=\rho_0$, corresponding to fixed passive cross-links (we consider timescales short compared to their lifetimes). 

\paragraph{Discrete variable for active bonds}
We treat the dynamics of the motors explicitly by introducing a discrete occupation variable $n_i$ for each bond $i$ such that
\begin{equation}
n_i=\Big\{
\begin{array}{cc}
1&{\text{if bond $i$ active (motor bound)}}\\
0&{\text{if bond $i$ passive (motor unbound)}}
\end{array}
\nonumber
\end{equation}
with stochastic switching between states $n=0$ and $n=1$ and the probability of being in state $n=l$ being $\ds P_l (t) = \aver{\delta_{n l}}_n$ with dynamics;
\beqa
\partial_t P_1 = - k_u P_1 + k_b P_0 \quad ; \quad P_0= 1-P_1\quad , 
\label{eq:b-u}
\eeqa
where $k_b$ and $k_u$ are the binding and unbinding rates respectively. 
We ignore the effect of mechanical forces on $k_b$ and $k_u$ but note that the stochastic dynamics of the bonds will depend on whether a motor is bound or not.

The micro-states of our system are then the set of displacements $\{ x_i\}$ and motor occupation numbers $\{ n_i \}$ of the bonds $i \in \{ 1 , \ldots,N\}$. The probability of finding the system in a particular micro-state is denoted $P(\{ x_i\} ,\{ n_i\},t)$. If  the dynamics of $n_i$ (binding/unbinding of motors) is the slowest process, i.e. fluctuations in  $x_i$, relax  faster, then 
$P(\{ x_i\} ,\{ n_i\},t)=P(\{ x_i\} | \{ n_i\};t) P (\{ n_i\},t)$, 
where $P(\{ x_i\} | \{ n_i\};t)$ is the {\em conditional probability} of finding a set of $\{x_i\}$, {\em given} the set of occupation numbers $\{ n_i\}$. We can describe the dynamics $P (\{ n_i\},t)$ using  $N$ uncoupled copies of eqn.~(\ref{eq:b-u}) while 
the equation of motion for $P(\{ x_i\} | \{ n_i\},t)$ is 
\beqa
\ds \partial_t P(\{ x_i\} | \{ n_i\}) = - \sum_{j=i}^N  {\partial J_j  \over \partial x_j} \label{eq:smol}
\eeqa
with $\ds J_j  = - D{\partial \over \partial x_i} P(\{ x_i\} | \{ n_i\}) + P(\{ x_i\} | \{ n_i\}) {g_j \over \xi}$
where $D$ is the amplitude of the displacement fluctuations and $\xi$ is a local friction. 
The current $J_j$ has a deterministic contribution from fluxes due to the forces, $g_j$, and a contribution from the fluctuations ($\propto D$).
The forces, $g_j$, have an elastic and active part: 
\bem\ds
g_j (\{ x_i\} , \{ n_i\}) = -  {\partial \over \partial x_j} H_{\text{elastic}} - n_j f_m
\nonumber
\eem
~where $f_m$ is the contractile active force exerted by a motor and the elastic energy, 
\bem
H_{\text{elastic}} = \sum_{i=1}^N \frac{1}{2} k^{\rm{eff}} x_i^2 \, , 
\nonumber
\eem
~with $k^{\rm{eff}}=(\phi_1\frac{k}{2}+\phi_2k)=\rho_0\frac{k}{2}$ the effective spring constant ($\frac{k}{2}$ is the spring constant of a single bond~\footnote{We could include nonlinear stiffening effects of activity on the elasticity of the individual elements here by having $k^{\mbox{eff}} = k_0 + \Delta \mu k_1$~\cite{Liverpool2009}.}).
The steady state distribution can then be calculated as 
\bem
P_{ss}(\{ x_i\} | \{ n_i\}) = {1 \over {\cal{Z}} (\{ n_i\}) } \exp \left[ - \beta H \right] \, , \; 
\nonumber
\eem
where 
$\beta^{-1} = D \xi \ne {k_B T}$ 
and 
\bem
{\cal{Z}} (\{ n_i\}) = \int \prod_{i=1}^N dx_i \; P_{ss} (\{ x_i\} | \{ n_i\}) \; 
\nonumber
\eem
with 
\beq
H (\{ x_i , n_i \}) = H_{\text{elastic}} + \sum_{i=1}^N f_m n_i x_i  \; . 
\nonumber
\eeq
Averages of physical quantities are as usual:
\bem B(t) = \aver{B(\{ x_i\})} = \int \prod_{i=1}^N dx_i \; B(\{ x_i\}) \; P (\{ x_i\} | \{ n_i\},t) \, , \; \nonumber
\eem
and in the steady-state they can be obtained from derivatives of the generating functional $ \ln {\cal{Z}}(\{n_i\})$ which must be averaged over $\{ n_i \}$.

\paragraph{Mechanical response:} We now consider the mechanical response of the system to a macroscopic deformation $L_0 \rightarrow L_0 + \Delta L$, with $\Delta L = N b_0 U$, thus defining a macroscopic strain, $U$. 
Hence  $U=\frac{1}{Nb_0}\sum_{i=1}^Nx_i$ 
\footnote{This argument can be made local if $U$ varies in space over length scales much larger than the lattice spacing $b_0$.}
and  the relevant generating functional, 
\begin{align}
{\cal{Z}}(\{n_i\},U)=&\int \prod_i \dd x_i\,e^{-\beta H}\delta({\scriptstyle{\frac{\sum_i^Nx_i}{Nb_0}}}-U)
\label{eq:ZSQ}
\end{align}
We consider the evolution of the macroscopic stress \bem\sigma = \aver{\frac{1}{AN}\sum_{j=1}^N g_j }_{n_i,x_i} \eem with {\em time} 
after a step-strain, $U$.

In performing averages over $\{n_i\}$, 
we consider two limits determined by the timescales $k_u^{-1},k_b^{-1}$.

(i) For  $ t \ll k_b^{-1}$, $k_u^{-1}$, the motors are frozen in a particular set of $\{ n_i\}$, i.e. $n_i$ are quenched. Macroscopic quantities 
 can be calculated from derivatives of  $\ln {\cal Z} (\{ n_i \})$ averaged over $\{ n_i\}$, ${\cal{F}}_{\rm{qu}}=-\frac{1}{\beta}\aver{ \ln {\cal Z} (\{ n_i \} ) }_{n_i}$. 
The average is done 
over the initial distribution of occupation numbers, $P(\{n_i\}, t=0)$,
taken to be the steady-state distribution, $P_{ss}(\{n_i\})$ given in \footnote{Supplemental Materials\label{SI}}.
This gives the stress: 
\begin{align}
 \sigma^{\rm{qu}}&=\frac{1}{V}\frac{\dd{\cal{F}}_{\rm{qu}}}{\dd U}=\frac{Nb_0}{V}\Big (k^{\rm{eff}}b_0U+a_0f_m\Big )\label{eq:stressquench}
\end{align}
where $a_0=\frac{1}{2}m_0(1-p_0^2)(\rho_0-1)=\frac{k_b}{k_u + k_b}$ is the mean fraction of active bonds. The ground state displacement $\tilde{x}_0=b_0U_0=-\frac{a_0f_m}{k^{\rm{eff}}}$ (where $U_0$ is the strain for which $\sigma=0$) and  elastic constant $\tilde{E}=\frac{V}{N^2b_0^2}\frac{\dd\sigma}{\dd U}=\frac{k^{\rm{eff}}}{N}$.
This is equivalent to  a `mean
field model'  in which $n_i$ is fixed to its average  value  $a_0$.

(ii) For $t \gg k_u^{-1},k_b^{-1}$, the motor variable, $n_i$, is annealed. We can average over $\{n_i\}$  in eqn. (\ref{eq:ZSQ}) and hence obtain averaged quantities from derivatives of ${\cal{F}}_{\rm{an}}=-\frac{1}{\beta} \ln \aver{{\cal{Z}}(\{n_i\})}_{n_i}$ also averaged over $P_{ss}(\{n_i\}) $. 
The stress, $\sigma^{{\rm{an}}}=\frac{1}{V}{\frac{\dd{\cal{F}}_{\rm{an}}}{\dd U}}$, is given by (see \footnotemark[\value{footnote}] for details):
\be
\sigma^{{\rm{an}}}
\approx \frac{Nb_0}{V}\Big ((k^{\rm{eff}}-a_0(1-a_0)\beta
    f_m^2)b_0U+a_0f_m\Big ),\nonumber
\fin
leading to the ground state local displacement $\tilde{x}_0^{\rm{an}}=-\frac{f_ma_0}{k^{\rm{eff}}-a_0(1-a_0)\beta
    f_m^2} < \tilde{x}_0^{\rm{qu}}$
and elastic constant, $\tilde{E}^{\rm{an}}=\frac{1}{N}\left(k^{\rm{eff}}-a_0(1-a_0)\beta f_m^2\right)< \tilde{E}^{\rm{qu}}$ 
up to $O(f_m^2)$ \footnote{Note that the second term is small compared to the first since, for linear elasticity, $\frac{\beta f_m^2}{k^{\rm{eff}}}\sim\beta k^{\rm{eff}}x^2\ll 1$.}. 
These results are equivalent to taking $n_i=a$, where $a$ fluctuates with a Gaussian distribution;
$P(a)=\frac{1}{\sqrt{2\pi \nu_{a}^2}}\; e^{-\frac{(a-a_0)^2}{2\nu_{a}^2}}\label{eq:P(m_p)}$
with mean $a_0$ and variance $\nu_{a}^2=a_0(1-a_0)/N$,
where finite $\nu_a$ could reflect fluctuations in the polarity
$p$ as well as number of motors $m$. 

\begin{figure}[hbt]
\begin{centering}
\includegraphics[width=0.5\textwidth]{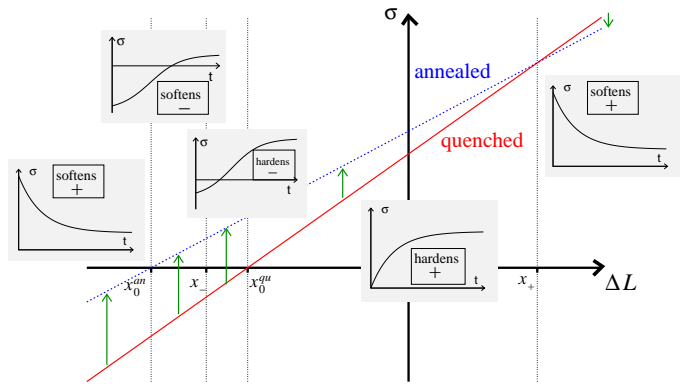}
\caption{Graph of the restoring stress, $\sigma$, versus the applied deformation, $\Delta L$. Arrows indicate direction of stress evolution. Grey insets: schematic drawings of $\sigma(t)$ showing decreasing/increasing $|\sigma|$ over time (softens/hardens) where $-$ indicates $\sigma$ changes sign.}
\label{fig:punch}
\end{centering}
\end{figure}

Our results imply a rich variety of dynamic responses depending on the amount of deformation. 
Initially after a deformation, the material will respond with effective elastic properties given by the quenched motor variables but on longer timescales 
with that of the annealed motor variables. 
Over time the restoring stress 
can increase, decrease or even change sign as  $\sigma^{\rm{qu}}(U)\rightarrow \sigma^{\rm{an}}(U)$, depending on the amount of deformation applied.
This can lead to apparent stress hardening or softening with time. We define hardening/softening as a stress at long times that has a greater/lower magnitude than the initial stress. It is worth noting that the mechanical response of this material is naturally asymmetric, with stretch different  from compression. The classes of behaviour are schematically described in Fig.~\ref{fig:punch}.  It is useful to define $x_{+}$ as the deformation when $\sigma^{qu}(x_+) = \sigma^{an}(x_+)$ and 
$x_{-}$ where $\sigma^{qu}(x_-) = -\sigma^{an}(x_-)$.
For $\Delta L> x_+$, the system softens (stress decreases) and doesn't change sign (denoted + in Fig. \ref{fig:punch}). For $\tilde{x_0}^{qu} < \Delta L < x_+$, it hardens (stress increases) without changing sign (+).  For $\tilde{x_0}^{qu} > \Delta L > x_-$, the stress changes sign but the magnitude of the final stress is larger than that of the initial stress. This would be perceived as dynamic hardening with a change of sign.

\paragraph{The contractile active element}
Finally we outline a model for the mechanics of an individual active bond in more microscopic detail. It is the linear limit of a more complex model \cite{Liverpool2009}. As shown in Fig.~\ref{fig:1Dmuscopic}~inset, it is made up of a pair of antiparallel polar filaments linked by a motor cluster  (as parallel filaments lead to sliding without force generation). In the absence of forces the filaments have length $b_0$. The filaments are linearly elastic, with a segment of length $l$ having  spring constant $k(l) = k_0/l$ (springs in series).  The motor cluster is made up of two motors connected by an elastic element of stiffness $k_m$, which we consider first as much stiffer than the filaments, $k_m \gg k$.   The motors  `slide'  along the filaments towards the + end until the elastic forces in the cluster are equal to their stall force, $f_s$~\cite{alberts}. In this stationary 
state, the two motors apply equal and opposite forces $\pm f_s$  to the two filaments  at their respective attachment points which each split a filament into two segments of length $l_a$ and $ l_b=b_0-l_a$ with spring constants $k_{a,b}=k(l_{a,b})$. On average, the motor cluster is attached to the midpoint of the filaments, $l_a=b_0/2+s$. 
The force $f_s$ applied to the filaments leads to deformations $X_{a,b}$ of the two segments and a 
jump in the tensions $f_s = \tau_b-\tau_a$ between them
, where $\tau_{j}=k_{j} X_{j}$ 
are the tensions in the respective segments, $j=\{a,b\}$. 
Elasticity of the motor cluster implies $f_s/k_m =2s + X_a - X_b$.
When there is no external force applied ($\tau_a+\tau_b=0$), the total deformation, $X=X_a+X_b$,  is calculated as
$X=-b_0 \frac{f_s^2}{4 k_0^2}\left(1+\frac{2k_0}{b_0 k_m}\right)$. This is negative for all finite $k_0,k_m,f_s$ and therefore the active bond is
always contractile. If $k_0\rightarrow\infty$ (rigid filaments),
$X\rightarrow 0$ as expected. The contractile force in the model
above, $f_m=-k^{\rm{eff}}X$. 

Let us now estimate typical timescales and moduli for a cytoskeletal extract such as actin-crosslinkers-myosin-ATP% that motivated this study
~\cite{Gardel2003,Gardel2004,Mizuno2007}.
%We estimate typical timescales and moduli by mapping our calculation onto this system. % to verify if our assumptions are reasonable.
%\paragraph{Estimate values for an example system}
%We consider an actomyosin network as an example system. 
%We consider concentrations of actin and associated proteins  such that the 
Segments between cross-links of typical length $b \sim 0.1 \mu$m have a relaxation time $\tau_b \sim  {b^4 \eta \over k_B T l_p} \sim 10^{-6}$s
where $\eta\sim 10^{-3}\text{Pa}\,\text{s}$ is the viscosity of water and  $l_p \sim 15
\mu$m is the persistence length of
actin~\cite{IsambertMaggs96,Kroy1997a,Liverpool2001}. 
%The precise structure of the contractile motor elements is still not understood~\cite{Silva2011},  however if 
Assuming $\tau_m \sim 1\,$s~\cite{Veigel2003,Mizuno2007},
%they remain bound for 10's of seconds~\cite{Veigel2003,Mizuno2007}, $\tau_m \sim 10$s, 
then  $\tau_b \ll \tau_m$ as we have assumed above.
The stall force $f_s\sim f_m\sim 2\,{\text{pN}}$~\cite{Molloy1995},
 $k^{\rm{eff}}\sim k_BT l_p^2/b^4$~\cite{MacKintosh1995}, and $\beta^{-1}\sim k_BT/10$~\cite{Mizuno2007,Liverpool2003a} can similarly be estimated. %for a semiflexible filament.

\paragraph{Discussion}
We have developed and studied a microscopic stochastic model of a 1D disordered solid composed of elastic and active elements appropriate for timescales where the cross-links are fixed but the dynamics of motor (un)binding are important.
We find that due to the (un)binding of motors, this active solid is a highly responsive material with a variety of different mechanical responses depending sensitively on the amount of applied deformation.
This is due to the fact that the activity of motors leads to a contracted ground state and a modified elastic constant whose properties depend on the local (re)organisation of contractile elements. 
On timescales long compared to motor binding dynamics the contraction is greater and the material softer than 
on timescales shorter than the motor dynamics.  This 
results in a variety of counterintuitive mechanical behaviours e.g. initially after a deformation, the direction of elastic response can even be opposite in direction to that on long timescales~\cite{Wottowah2005}. This suggests  that the rich variety of behaviour shown by the cytoskeleton - its ability to adapt its properties to perform the mechanical tasks involved in  cell division or cell locomotion can be understood as a natural consequence of this type of collective dynamics.

While we have for reasons of clarity restricted ourselves to a one dimensional model, it  
is natural to consider higher dimensions, in which empty bonds can be included as long as the density $\rho$ remains above the percolation threshold for the chosen lattice. 
We note that if the density of filaments is reduced below the percolation threshold the system would appear to expand, since the network will break apart and lose contractility.
We have studied the system to lowest order in $\Delta \mu$:  to capture the behaviour at high activity  a number of other nonlinear,
  effects must be considered~\cite{Alvarado2013}.
Clearly including the force dependence of motor (un)binding rates $k_u ({x_i})$, $k_b({x_i})$, can can couple the rich behaviour we have described above to external mechanical cues and lead to 
mechanical analogues of switches and logic gates. 
Our framework can therefore be the starting point for more complex and realistic models 
%that can  be 
extended to  timescales comparable to dynamics of the system architecture (lifetime of passive cross-links) and start to approach quantitative models of whole cell %motile 
behaviour. 

\acknowledgments{We  would like to thank the Isaac Newton Institute for Mathematical Sciences (Cambridge) where some of the work was undertaken. This work was supported by EPSRC grant  EP/G026440/1.}

\bibliography{active}

\end{document}